# Inverse design of the transmission matrix in a random system using Reinforcement Learning

Yuhao Kang

Amazon, Seattle

yuhaok@uchicago.edu

Abstract

This work presents an approach to the inverse design of scattering systems by modifying the transmission matrix using reinforcement learning. We utilize Proximal Policy Optimization to navigate the highly non-convex landscape of the object function to achieve three types of transmission matrices: (1) Fixed-ratio power conversion and zero-transmission mode in rank-1 matrices, (2) exceptional points with degenerate eigenvalues and unidirectional mode conversion, and (3) uniform channel participation is enforced when transmission eigenvalues are degenerate.

Engineering wave propagation is a fast-moving domain. Singularities of the scattering matrix (SM), or sub-SM, such as the transmission matrix (TM) or reflection matrix (RM) encode the scattering behavior of an open system and can be exploited in sensing, switching, lasing and energy deposition [1,2]. Open systems can be described by effective non-Hermitian Hamiltonians, and their resonance frequencies corresponds to poles of SM. Frequency points at which the response vanishes are described by zeros, which are also usually complex value. In unitary systems, the SM displays complex conjugate pairs of poles and SM-zeros in the complex frequency domain, while the TM exhibits complex conjugate pairs of TM-zeros [3]. Incident radiation is completely absorbed when a zero of the SM is brought to the real axis. Such coherent perfect absorption (CPA) is the time reversal of an outgoing wave at the lasing threshold [4]. Similarly, transmissionless or reflectionless modes can be achieved when zeros of the TM or RM are brought to the real axis [3,5]. Exceptional points (EPs) indicate a unique class of singularities at which eigenvalues and eigenstates coalesce simultaneously. The EP of the effective Hamiltonian indicates the poles degenerate at a single frequency. The response of systems near EPs to perturbations is significantly enhanced, scaling with the n-th root of the perturbation strength for an n-th order EP [6]. The EP of the SM at a single frequency along with CPA can be used to realize a 50:50 constant power splitter [7].

While these phenomena have been widely discussed, precisely engineering random media with these singularities at the target frequency remains challenging. The complexity comes from three key factors: First, random media lack inherent symmetry and typically support multiple quasi-normal modes at a given frequency, making it impossible to simplify the analysis to two or three mode. Secondly, these singularity points exhibit extreme sensitivity to perturbations, and controlling their directional movement is hard. Finally, random system has many degrees of freedom that could be controlled in the design process. Determining which parameters to adjust and how to adjust them becomes complex. Our work demonstrates that Reinforcement Learning (RL) techniques offer an approach to solving these complex design problems.

RL for inverse design of complex scattering system and structure optimization is of growing interest. RL [8–11] is a machine learning method in which an agent learns to make decisions by interacting with an environment, receiving feedback based on rewards, and adjusting its policy to maximize the cumulative reward over time. Unlike supervised learning which requires labeled training data, RL algorithms learn through trial and error, making them particularly suitable for problems where the optimal solution path is unknown, but outcomes can be evaluated. Inverse designing a random system with a customized scattering property is a challenging task [12–17]. Due to the high degrees of freedom in the design space, and resonance and interference phenomenon, the objective function is usually highly oscillatory and non-convex. Small changes in design parameters can cause marked shifts in the objective function due to wave interference. Multiple local optima coexist so that optimization algorithms may get trapped in suboptimal solutions. Thus, finding global optima of an objective function generally requires a computation resource that grows exponentially relative to the complexity of the design space. In addition, the sampling process of a photonic system is computationally expensive. Differential-equation solvers, such as finite-difference time-domain (FDTD) or the finite element method (FEM) solver, demand significant time to solve Maxwell equation and simulate light propagation through complex structures.

RL is a good candidate for photonic design since its exploratory nature allows it to effectively navigate the highly non-convex landscapes of photonic systems. By iteratively exploring the design space and learning from the outcomes of its actions, a RL algorithm is more likely to discover global optima or an acceptable superior local optima. Auxiliary models can be used to replace the expensive wave simulation, for instance, spectral predictors [15,18,19] can be used to predict the spectrum without running actual simulation for each training step.

RL algorithms have been applied to various optical problems [20–26]. REINFORCE algorithm [27] is applied to design High-efficiency grating couplers, which achieved 86.8-87.8% coupling efficiency and has approached the theoretical maximum efficiency of 88% [25]. Deep Q-Network (DQN) [28] and Proximal Policy Optimization (PPO) [29] algorithms are used to design photonic crystal nanobeam cavities, which realize high Q-factors while keeping small modal volumes [22]. Double Deep Q-Learning Network (DDQN) [30] is used to increase transmission efficiency of metasurface holograms [24] and design all-dielectric reflective color filters, which can achieve pure red, green and blue colors [23].

In this work, we show that RL can be used to manipulate the scattering property of a random system. Without loss of generality, we will use TM as the target in this work. There are three types of values that are of interesting: 1) The frequency of the transmission zero [3]. The determinant of transmission matrix can be expresses as a rational function based on the Heidelberg model [31–33],

$$\det(T(f)) \propto \frac{\prod_{i=1}^{M-N}(f-\eta_i)}{\prod_{j=1}^{M}(f-p_i)}, \qquad (1)$$

where M and N indicates number of resonant modes and propogating channels that dominate the transmission at a given frequncy $f$. $p_i$ indicates the poles. At the complex frequencies $\eta_i$, we can adjust the input pattern to produce zero transmission. 2) Eigenvalue. For a transmission matrix T, eigenvalues λ satisfy $T|v\rangle = \lambda|v\rangle$, where $v$ is an eigenvector. Eigenvalues describe how the system transforms incoming electromagnetic fields into outgoing fields while preserving the field patterns. 3) The transmission eigenvalue is derived from the singular value decomposition (SVD) of the TM $T = U\Sigma V^\dagger$, where Σ is a diagonal matrix with singular values $\Sigma = \text{diag}(\sqrt{\tau_1}, ... \sqrt{\tau_N})$, and the transmission eigenvalues $\tau$ are the square of singular values.

In the following section, we will show how to (1) construct a rank-1 TM and move the zero-transmission frequency $\eta$ to the target frequency, (2) make the eigenvalue λ degenerate at the target frequency, or (3) make the transmission eigenvalue $\tau$ degenerate at target frequency.

Our system is a billiard with perfect electrically conducting walls with two inputs and two outputs, which contains at most 20 circlular scatterers. The scatterers have a higher dielectric constant ε=2.1 than the background of air. The source frequency is 15 GHz. Perfectly matched layer (PML) absorbing boundaries are at the ends of the waveguides. The Meep FDTD solver [34] is used to calculate the field distribution. The scatterers can be positioned to control the transmission spectrum and achieve specific properties in the TM.

The process of photonic design can be mapped to a sequence of interdependent decisions, at which each action, e.g., adjusting the position of scatterers, influences subsequent states and ultimately the final performance. RL is specifically designed for such sequential decision-making problems, making it a natural fit for iterative design processes.

The state is fully characterized by the normalized spatial coordinates of all scatterers. For a system with n scatterers, the state vector is represented as $[x_1, y_1, x_2, y_2, ..., x_n, y_n]$, where each coordinate pair $(x_i, y_i)$ specifies the position of the i-th scatterer. The x and y coordinates of all scatterers are normalized to the range [-1, 1]. The action space indicates incremental adjustments to each scatterer's position, formatted identically to the state vector: $[\Delta x_1, \Delta y_1, \Delta x_2, \Delta y_2, ..., \Delta x_n, \Delta y_n]$. These movements are also normalized to [-1, 1] and subsequently scaled by a small factor 0.005 to facilitate gradual, controlled optimization of the scatterer arrangement. The Policy indicates the probability distributions over possible position adjustments given a system, which guides the motion of the scatterers to maximize the episode reward. When position adjustments lead to two scatterers overlapping, only one scatterer effectively remains in the simulation. This implementation follows the "last object wins" principle in Meep solver, where later objects take precedence when overlaps occur. This mechanism introduces an additional degree of freedom to the system - the actual number of scatterers can change throughout the optimization process and make the design space more flexible.

Here, we choose the PPO architecture, an effective on-policy RL algorithm as shown in Fig. 1b. The input state, which is the set of normalized scatterer coordinates, first passes through a flattening extractor which processes and prepares the data for parallel evaluation. Then the data is propagated into two specialized neural pathways: a policy network (Fig. 1b left) and a value network (Fig. 1b right), each comprised of two fully connected Linear layers along with ReLU activation functions to capture the nonlinear relationships. The policy network then passes the value to another Linear layer that outputs a probability distribution over possible scatterer adjustments, while the value network's Linear layer estimates the expected cumulative rewards for the current configuration. This structure enables PPO to simultaneously learn the optimal positioning strategies and evaluate the state values, maintaining stability through its clipped objective function that prevents excessive policy updates during training [29].

The cost function is used to describe the distance between the target TM and the actual simulation. One should design a scale-free cost function, i.e., if the TM is multiplied by any scalar, the cost should remain unchanged. Otherwise, RL may drive the overall matrix scale toward zero.

The cost functions used to realized various targets is listed below.

| Target | Cost function |
| --- | --- |
| N*N Rank-1 TM | $$T = U \begin{bmatrix} \sqrt{\tau_1} & & \\ & \ddots & \\ & & \sqrt{\tau_N} \end{bmatrix} V^\dagger$$ $$\text{cost} = 1 - \frac{\tau_1}{\sum_{i=1}^{N} \tau_i}$$ |
| 2*2 TM with Degenerate Eigenvalue | $$T = \begin{bmatrix} t_{11} & t_{12} \\ t_{21} & t_{22} \end{bmatrix}$$ $$\lambda_\pm = \frac{1}{2}[(t_{11}+t_{22}) \pm \sqrt{(t_{11}-t_{22})^2 + 4t_{12}t_{21}}]$$ $$\text{cost} = \frac{\left|(t_{11}-t_{22})^2 + 4t_{12}t_{21}\right|}{\varepsilon + \sum_{i=1\&2, j=1\&2} |t_{ij}|^2}$$ |
| 2*2 TM with degenerate transmission eigenvalue | $$\text{cost} = \frac{\left|\sqrt{\tau_1} - \sqrt{\tau_2}\right|}{\varepsilon + \sqrt{\sum_{i=1\&2, j=1\&2} |t_{ij}|^2}}$$ |

Table 1. **Cost function for TM targets.** $\varepsilon = 10^{-8}$ at denominator to avoid divergence.

At each step t, the Reward is the negative of the cost function, $R_t = -\text{cost}$. The episode reward is the sum of all individual rewards received across each timestep of an episode, $R_{episode} = \sum_{t=0 \sim L-1} \gamma^t R_t$, where $\gamma = 0.999$, a discount factor for future rewards, determining how much the agent values long-term rewards. A high discount factor encourages the agent to consider long-term rewards, which aligns with the goal of finding a stable, optimal configuration over multiple steps. L=1024 indicates the maximum length of a single episode. The target of RL is to maximize episode reward $R_{episode}$.

Our photonic inverse design framework implements a cyclical process of sampling, learning, and resetting. The 1st episode starts with random initialization. During sampling, the agent observes the current normalized positions of all scatterers and guided by its policy network, outputs a probability distribution over possible position adjustment. Actions are sampled from this distribution and applied as small perturbations to each scatterer's coordinates. For each adjustment, we use the Meep FDTD solver to calculate the characteristics of the TM and the immediate reward. The episode ends either by reaching the termination condition where the cost function reach the expected threshold or exhausting the maximum 1024 steps. The environment is then reset according to our hybrid strategy: with 70% probability, we initialize the next episode with small Gaussian perturbations around the best-known configuration; otherwise, we generate an entirely new random scatterer positions to facilitate global exploration. This reset mechanism prevents convergence to suboptimal designs while efficiently exploiting promising regions of the design space. We do not use any auxiliary model to predict the TM. The main target of this work is to manipulate the singularities of the TM. The property near the singularity is usually highly sensitive, and auxiliary model does not have enough accuracy.

The ideology of PPO is to continuously explore the design space while gradually shifting its exploration toward more promising regions. PPO uses a stochastic policy that introduces randomness in actions. The policy is being updated to increase the probability of actions that lead to higher rewards, shifting the distribution of rewards upward. While individual outcomes remain random, their statistical distribution improves over time so that the individual structure has greater opportunity to reach the cost threshold. Figure 1c shows a typical training process [35]. The upper panel is the reward of individual step, which appear random and reflect the stochastic nature of individual scatterer adjustments in the complex electromagnetic landscape. The episode reward shown in the lower panel increases gradually, which indicates the average performance across many attempts is improving. PPO is effectively lifting the baseline of performance.

In the following section, we show multiple converged results obtained though PPO based on the various cost functions in Table 1.

**Rank-1 TM.** It is useful to develop a power converter that can produce multiple outputs with fixed power ratios and fixed phase difference, regardless of the shape of input wavefront. When a TM has rank 1, it can be expressed as an outer product of two vectors $T = |v_{out}\rangle\langle v_{in}|$, where $v_{out}$ decides the fixed output distribution pattern and $v_{in}$ is a general vector. For any arbitrary input wavefront $u_{in}$, the output field is simply $v_{out}$ multiplied by a scalar coefficient, $T|u_{in}\rangle = |v_{out}\rangle\langle v_{in}|u_{in}\rangle$. When $u_{in}$ is orthorgonal to $v_{in}$, transmission vanishes.

Figure 2a shows the speckle pattern due to individual inputs are excited in a rank-1 TM system. The output field ratio is the same no matter the combination of the top and bottom incoming signal.

Since N*N rank-1 matrix has (N-1) zero-transmission eigenchannels, in our case, the second transmission eigenvalue vanishes, as shown in Fig.2b. The existence of the transmission-less mode indicates that the target frequency is a transmission zero for the TM. Figure 2c shows the phase map of det(T), with phase increasing in counterclockwise direction by $2\pi$ around this transmission zero point [3].

An N*N rank-1 matrix has (N-1) zero eigenvalues, with the remaining eigenvalue either zero or non-zero. At the transmission-zero point, det(T) is a bridge that connects the eigenvalue and transmission-zero frequency. Since the product of all eigenvalues $\lambda_j$ is det(T):

$$\det(T(f)) = \prod_{j=1}^{n} \lambda_j(f) \qquad (2)$$

Equations 1 and 2 describe the TM from different perspectives. Equation 1 describes the interference of modes in the complex frequency plane, while Eq.2 describes the spatial domain at a single frequency. At a transmission-zero frequency $\eta$, $\det(T(\eta)) = 0$, and at least one eigenvalue $\lambda_0(\eta) = 0$. When TM is doubly degenerate at the transmission zero $\eta = \eta_1 = \eta_2$, Eq. (1) indicates $\det(T(f)) \propto (f - \eta)^2$. Considering the Taylor expansion of eigenvalues near $\eta$, there are two cases:

a) $\lambda_0(f) \sim (f - \eta)$ and $\lambda_1(f) \sim (f - \eta)$, so that there are double zero eigenvalues at $\eta$, ie. $\lambda_0(\eta) = \lambda_1(\eta) = 0$. b) $\lambda_0(f) \sim (f - \eta)^2$, then there is only single zero eigenvalues $\lambda_0(\eta) = 0$. We can conclude the degeneracy of transmission frequency zero is higher or equal than the degeneracy of eigenvalue zero.

To realize a second-order transmission zero frequency, a natural thought is to construct a TM with double degenerate zero eigenvalues. According to the analysis above, the degeneracy of transmission zero frequency must be two and the target frequency would be an EP of transmission zeros. However, our experience shows this approach is not feasible. The theoretical framework depends on achieving exact mathematical zeros at the target frequency. In practice, even when we successfully create a 2×2 transmission matrix with two small eigenvalues, these values never reach mathematically zero and we verified the phase winding number around the target frequency is 0 rather than $4\pi$ in those configurations. The root cause is due to probability: double zero eigenvalues represent only one pathway (case a) to achieve degenerate transmission zeros, while the system has higher probability with fewer eigenvalues approaching zero with higher order (case b). In short, double zero eigenvalues is a sufficient but not necessary condition for degenerate transmission zero frequencies, making this specific configuration statistically rare and hard to realize.

**Degenerate eigenvalues.** When matrix has degenerate eigenvalues, its corresponding eigenvector will coalesce around that degenerate point. For N×N matrix with normalized eigenvector $|v_{1\sim N}\rangle$, we can define Eigenvector Coalescence $C = 1 - \prod_{i<j}[1-\langle v_i|v_j\rangle]^{\frac{2}{N(N-1)}}$. When N=2, $C = \langle v_1|v_2\rangle$.

We construct a system with a TM with two degenerate eigenvalues. Fig.3a shows the coalescence map near the target frequency and perfect coalescence of 1 at the target point. In the phase map, the location perturbation of a single scatterer is introduced as another dimension.

Given mode conversion rate $C_{a\to b} = |\langle v_b|T|v_a\rangle|^2$, we can define the asymmetry factor of the mode conversion rate between any two modes $|v_a\rangle$ and $|v_b\rangle$ as $A(v_a, v_b) = \frac{|C_{a\to b} - C_{b\to a}|}{C_{a\to b} + C_{b\to a}}$ [36].

The asymmetry factor A=1 indicates this is a unidirectional conversion from one mode to another one, and A=0 indicates the conversion rate between those two modes are the same. We will analyze the asymmetrical effect for two types of mode basis near the EP.

Type-1 mode is Jordan canonical form with one eigenvector $|v_{EP}\rangle$ and one generalized eigenvector $|v_g\rangle$. When 2*2 TM has two degenerate eigenvalues $\lambda_0$, TM can be diagonalized to Jordan format.

$J_0 = P^{-1}T_{EP}P = \lambda_0 I + \begin{bmatrix} 0 & 1 \\ 0 & 0 \end{bmatrix}$ where $P = \begin{bmatrix} v_{EP} & v_g \end{bmatrix}$, $v_{EP}$ is the eigenvector, $|v_g\rangle$ is the generalized eigenvector which satisfy

$$T_{EP}|v_{EP}\rangle = \lambda_0|v_{EP}\rangle$$

$$T_{EP}|v_g\rangle = \lambda_0|v_g\rangle + |v_{EP}\rangle \qquad (3)$$

Eq.3 indicates an asymmetrical mode conversion, i.e., $|v_{EP}\rangle$ can only be transmitted to itself, while $|v_g\rangle$ can be transmitted as superposition of $|v_{EP}\rangle$ and $|v_g\rangle$. $|v_g\rangle$ and $|v_{EP}\rangle$ form complete basis of input and output profile. $A(|v_{EP}\rangle, |v_g\rangle) = 1$ near EP, as shown in Fig. 3b. The conversion between Type-1 mode in our system is shown in Fig.3c.

Type-2 mode is the standard eigenvectors basis $|v_{\pm}\rangle$ of the transmission matrix T, where $T|v_{\pm}\rangle = \lambda_{\pm}|v_{\pm}\rangle$. This basis is ill-defined at EP.

Considering the perturbation near the EP, $J = J_0 + \varepsilon \begin{bmatrix} w_{11} & w_{12} \\ w_{21} & w_{22} \end{bmatrix}$, where $\varepsilon$ is an infinitely small perturbation term. The eigenvalue can be expressed as $\lambda_\pm = \lambda_0 \pm \sqrt{\varepsilon w_{21}} + O(\varepsilon)$, thus,

$$A(|v_+\rangle, |v_-\rangle) = \frac{|\lambda_-|^2 - |\lambda_+|^2}{|\lambda_-|^2 + |\lambda_+|^2} = 2\left|\text{Re}(\frac{\sqrt{w_{21}}}{\lambda_0})\right|\sqrt{\varepsilon}.$$

At EP, the asymmetry factor A=0 since |v₊> and |v₋> coalesce (Fig. 3d). Away from EP, we expect the asymmetry factor A of those type-2 modes increase relative to perturbation with exponent 0.5. The fitting in Fig. 3e reveals the actual exponent of A relative to the scatter shift is 0.9. This mismatch is reasonable because the perturbation term $\varepsilon$ used in above derivation is imposed on TM, while the scatterer perturbation term does not have linear impact on TM perturbation.

**Degenerate Transmission eigenvalues.** Since $T = U\Sigma V^\dagger$, the transmitted power for a normalized input pattern $|\varphi\rangle$ can be expressed as

$P = \||T|\varphi\rangle\|^2 = \langle\varphi|V\Sigma^2 V^\dagger|\varphi\rangle = \sum_{ij}\langle\varphi|v_i\rangle\langle v_i|V\Sigma^2 V^\dagger|v_j\rangle\langle v_j|\varphi\rangle = \sum_i |\langle v_i|\varphi\rangle|^2 \tau_i$, where $v_i$ is the i-th column of V.

When transmission eigenvalues are degenerate at value $\tau$, $P = \tau \sum_i |\langle v_i|\varphi\rangle|^2 = \tau$. No matter how to change the input speckle pattern, the transmission is a constant $\tau$. TM can be expressed as a multiple of a unitary matrix, $T = \sqrt{\tau} U$.

Eigenchannel participation number $N_{eff} = (\sum_{n=1}^N \tau_n)^2 / \sum_{n=1}^N \tau_n^2$ range from 1 to N [37]. $N_{eff} = 1$ indicates one eigenchannel dominates the transmission, and $N_{eff} = N$ indicates all channels have the degenerate transmission. Fig. 4a indicates the degeneracy of transmission eigenvalue at 15 GHz. The variation of $N_{eff}$ relative to frequency and perturbation is shown in Fig. 4b, which oscillate between 1 and 2.

In this work, we use RL to achieve precise control over transmission matrix properties. By implementing the PPO framework that optimizes scatterer positions within a 2D billiard cavity, we successfully designed systems exhibiting rank-1 TM, degenerate eigenvalues, and degenerate transmission eigenvalues. The rank-1 TM enables creating systems that produce fixed output speckle pattern regardless of input speckle pattern, which can be used for fixed-ratio power splitting and beam shaping in optical systems. The rank-1 TM also indicates the target frequency is a transmission zero frequency, and transmission-less mode can be realized with fined-tuned incoming pattern, which has applications in optical filtering and selective transmission. These results illustrate how RL can overcome the challenges in photonic inverse design. While individual adjustments may appear chaotic, the RL agent is systematically learning to make statistically better decisions over time, gradually shifting the probability distribution of outcomes toward higher-quality solutions. We believe the framework here offers a general methodology for inverse design that can be extended to a wide range of problems. For instance, if the TM is replaced by the matrix relating the field channels inside a medium to the incident channels [38], this framework can be extended to design a system which can focus light to a target interior region. The phase derivative of det(T) is proportional to the density of states, which is the sum of Lorentzian of quasi-normal modes [39]. By specifying a target transmission time spectrum, the system can be inverse designed to possess quasi-normal modes with given frequencies, linewidths, and modal overlapping. This work creates a base solution that can be adapted for practical implementation and paves the way of experiment realization to manipulate TM in a random system, which opens new frontiers for wave filtering, constant-ratio power converter, and unidirectional wave conversion.

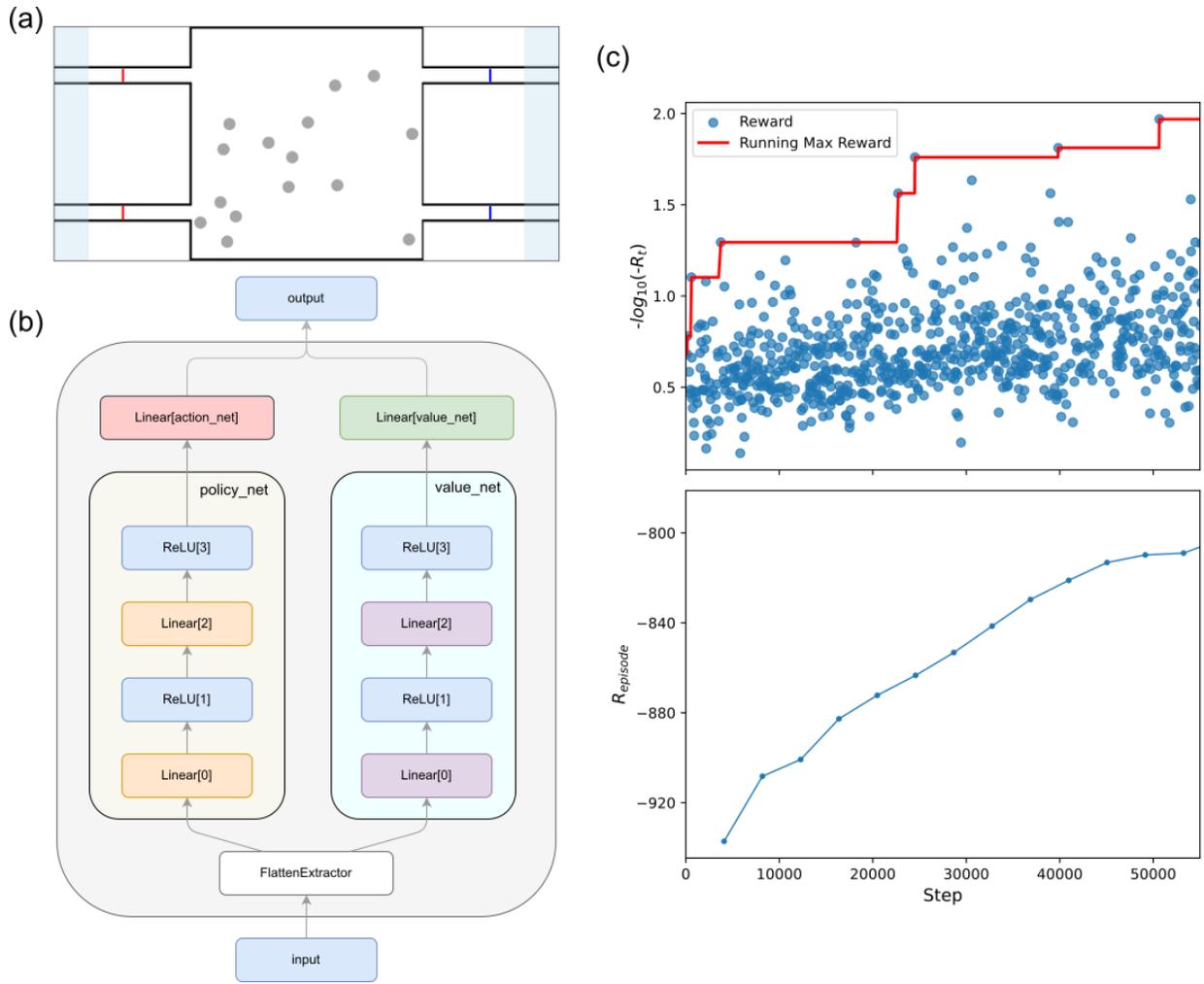

Fig. 1 **Random system and PPO**. a) A 2D billiard system with scatterers. Waveguide width is 1.2cm, length and width of billiard are both 20cm. The waveguide center distance to billiard horizontal midline is 6cm. Scatterer radius is 0.5cm. Red/blue line indicate the position of sources/monitors with frequency 15GHz. Light blue region indicates PML that absorb outgoing wave. b) PPO neural network architecture with dual pathways. Policy network (left branch) maps states to action probabilities. Value network (right branch) estimates state value function. c) Learning curve. Upper panel is the log scale of Reward at individual step; we plot the maximum Reward for every 128 steps. Red line indicates the running max of reward. The y axis scale is in log scale. Since Reward is always negative here, we use $-\log_{10}(-\text{Reward})$. The lower panel shows the cumulative episode rewards $R_{episode}$ with consistent improvement.

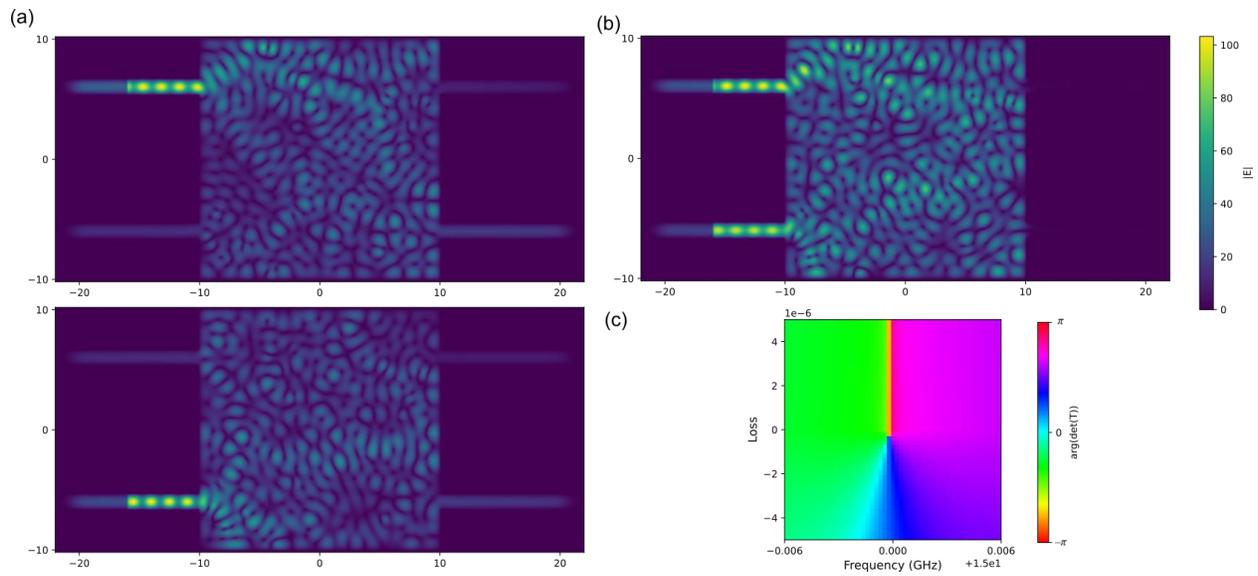

Fig. 2 **Rank-1 TM and zero-transmission mode**. a) Top panel: Excitation from top input waveguide only. Bottom panel: Excitation from bottom input waveguide only. There is interference pattern near source due to the interference between incoming and reflected wave. Both panels show consistent output field ratios between output waveguides despite different input excitation. The phase difference between bottom and top output is -2.7. The intensity ratio between bottom and top output is 4.7. b) Lowest transmission eigenchannel, with actual transmission eigenvalue of $1.06 \times 10^{-7}$. c) Phase map of det(T) near 15GHz. Loss/gain is at the lower/upper half plane. There is $2\pi$ phase winding around the zero-transmission frequency.

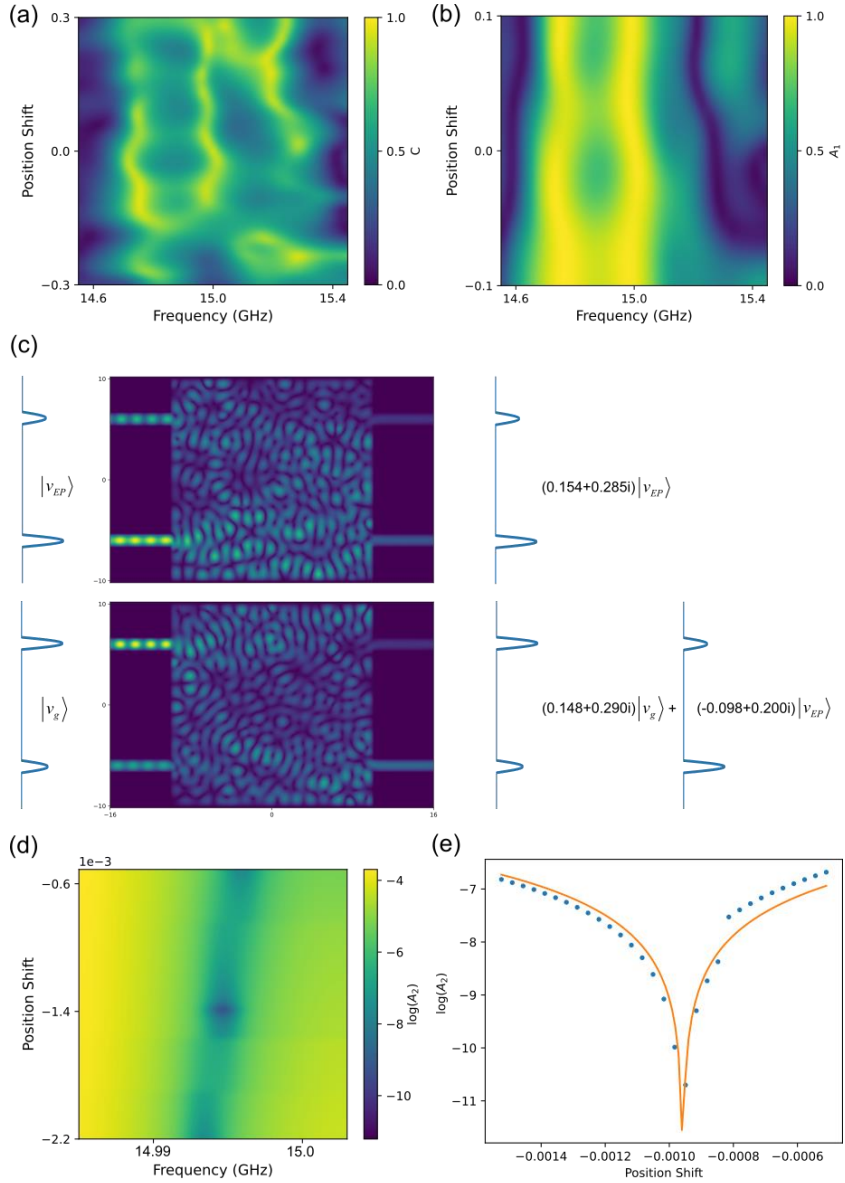

Fig. 3 **Degenerate eigenvalue and mode conversion**. a) At Freq=15GHz without perturbation, the eigenvector coalescence C=1. Away from the degenerate point, coalescence go down to 0. b) Type-1 Mode conversion asymmetry between $|v_{EP}\rangle$ and $|v_g\rangle$. c) The speckle patterns with normalized input $|v_{EP}\rangle$ and $|v_g\rangle$, respectively. $|v_g\rangle$ is choosen to be orthogonal to $|v_{EP}\rangle$. The output of $|v_{EP}\rangle$ excitation is still $|v_{EP}\rangle$ mode, while the output of $|v_g\rangle$ excitation is a superposition of $|v_g\rangle$ and $|v_{EP}\rangle$ mode. d) Type-2 Mode conversion asymmetry between $|v_+\rangle$ and $|v_-\rangle$. e) Fitting of d) near the dip at 14.995GHz. The fitted exponent in red line is 0.9, which deviates from 0.5 in perturbation theory.

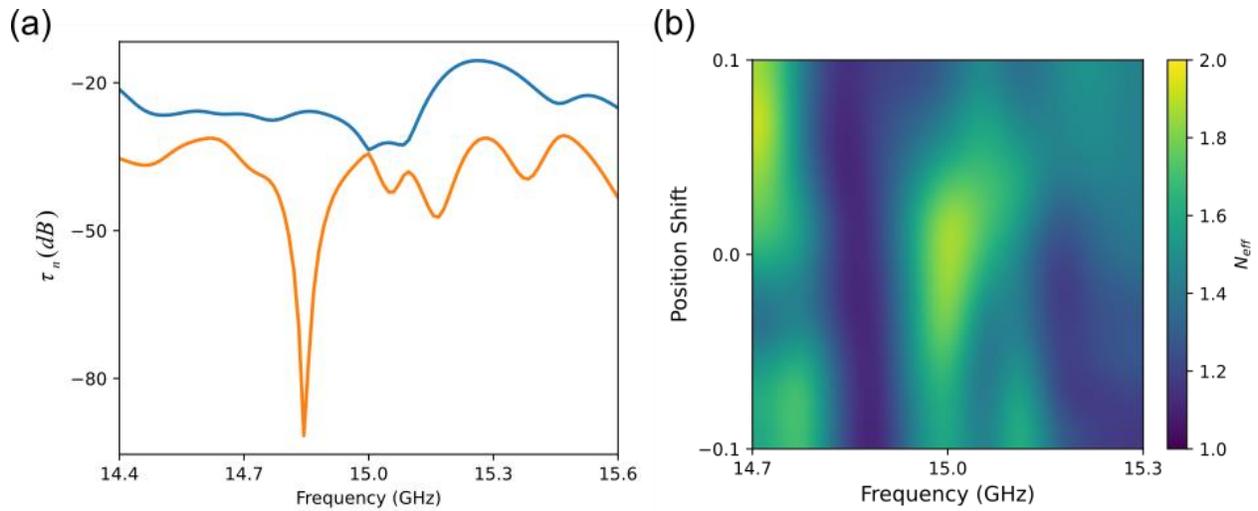

Fig. 4 **Degenerate transmission eigenvalues**. a) degenerate transmission eigenvalue at 15GHz. b) $N_{eff}=2$ at 15GHz and no perturbation. Away from the central frequency or adding more perturbation, $N_{eff}$ oscillates between 1 and N=2.

Code Availability

The code of this work is open sourced at https://github.com/yuhaokan/InverseDesignTM

Acknowledgements

The author thanks Prof. Azriel Z. Genack for valuable discussion and polishing that helped refine this manuscript.

References

[1] S. Longhi, $\mathcal{PT}$-symmetric laser absorber, Phys. Rev. A **82**, 031801 (2010).
[2] M.-A. Miri and A. Alù, Exceptional points in optics and photonics, Science **363**, eaar7709 (2019).
[3] Y. Kang and A. Z. Genack, Transmission zeros with topological symmetry in complex systems, Phys. Rev. B **103**, L100201 (2021).

[4] Y. D. Chong, L. Ge, H. Cao, and A. D. Stone, Coherent Perfect Absorbers: Time-Reversed Lasers, Phys. Rev. Lett. **105**, 053901 (2010).

[5] W. R. Sweeney, C. W. Hsu, and A. D. Stone, Theory of reflectionless scattering modes, Phys. Rev. A **102**, 063511 (2020).

[6] H. Hodaei, A. U. Hassan, S. Wittek, H. Garcia-Gracia, R. El-Ganainy, D. N. Christodoulides, and M. Khajavikhan, Enhanced sensitivity at higher-order exceptional points, Nature **548**, 187 (2017).

[7] J. Erb, N. Shaibe, R. Calvo, D. P. Lathrop, T. M. Antonsen, T. Kottos, and S. M. Anlage, Topology and manipulation of scattering singularities in complex non-Hermitian systems: Two-channel case, Phys. Rev. Res. **7**, 023090 (2025).

[8] D. Silver et al., Mastering the game of Go without human knowledge, Nature **550**, 354 (2017).

[9] R. S. Sutton and A. G. Barto, *Reinforcement Learning: An Introduction* (A Bradford Book, Cambridge, MA, USA, 2018).

[10] F. AlMahamid and K. Grolinger, *Reinforcement Learning Algorithms: An Overview and Classification*, in *2021 IEEE Canadian Conference on Electrical and Computer Engineering (CCECE)* (2021), pp. 1–7.

[11] A. K. Shakya, G. Pillai, and S. Chakrabarty, Reinforcement learning algorithms: A brief survey, Expert Systems with Applications **231**, 120495 (2023).

[12] L. Deng, Y. Xu, and Y. Liu, Hybrid inverse design of photonic structures by combining optimization methods with neural networks, Photonics and Nanostructures - Fundamentals and Applications **52**, 101073 (2022).

[13] S. Molesky, Z. Lin, A. Y. Piggott, W. Jin, J. Vuckovic, and A. W. Rodriguez, Inverse design in nanophotonics, Nature Photon **12**, 659 (2018).

[14] A. Y. Piggott, J. Lu, K. G. Lagoudakis, J. Petykiewicz, T. M. Babinec, and J. Vučković, Inverse design and demonstration of a compact and broadband on-chip wavelength demultiplexer, Nature Photon **9**, 374 (2015).

[15] D. Liu, Y. Tan, E. Khoram, and Z. Yu, Training Deep Neural Networks for the Inverse Design of Nanophotonic Structures, ACS Photonics **5**, 1365 (2018).

[16] Z. Li, W. Liu, D. Ma, S. Yu, H. Cheng, D.-Y. Choi, J. Tian, and S. Chen, Inverse Design of Few-Layer Metasurfaces Empowered by the Matrix Theory of Multilayer Optics, Phys. Rev. Appl. **17**, 024008 (2022).

[17] W. Ji, J. Chang, H.-X. Xu, J. R. Gao, S. Gröblacher, H. P. Urbach, and A. J. L. Adam, Recent advances in metasurface design and quantum optics applications with machine learning, physics-informed neural networks, and topology optimization methods, Light Sci Appl **12**, 169 (2023).

[18] S. So, T. Badloe, J. Noh, J. Bravo-Abad, and J. Rho, Deep learning enabled inverse design in nanophotonics, Nanophotonics **9**, 1041 (2020).

[19] S. An et al., A Deep Learning Approach for Objective-Driven All-Dielectric Metasurface Design, ACS Photonics **6**, 3196 (2019).

[20] S. So, T. Badloe, J. Noh, J. Bravo-Abad, and J. Rho, Deep learning enabled inverse design in nanophotonics, Nanophotonics **9**, 1041 (2020).

[21] W. Ma, Z. Liu, Z. A. Kudyshev, A. Boltasseva, W. Cai, and Y. Liu, Deep learning for the design of photonic structures, Nat. Photonics **15**, 77 (2021).

[22] R. Li, C. Zhang, W. Xie, Y. Gong, F. Ding, H. Dai, Z. Chen, F. Yin, and Z. Zhang, Deep reinforcement learning empowers automated inverse design and optimization of photonic crystals for nanoscale laser cavities, Nanophotonics **12**, 319 (2023).

[23] I. Sajedian, T. Badloe, and J. Rho, Optimisation of colour generation from dielectric nanostructures using reinforcement learning, Opt. Express, OE **27**, 5874 (2019).

[24] I. Sajedian, H. Lee, and J. Rho, Double-deep Q-learning to increase the efficiency of metasurface holograms, Sci Rep **9**, 10899 (2019).

[25] S. Hooten, R. G. Beausoleil, and T. V. Vaerenbergh, Inverse design of grating couplers using the policy gradient method from reinforcement learning, Nanophotonics **10**, 3843 (2021).

[26] C. Sun, E. Kaiser, S. L. Brunton, and J. Nathan Kutz, Deep reinforcement learning for optical systems: A case study of mode-locked lasers, Mach. Learn.: Sci. Technol. **1**, 045013 (2020).

[27] R. J. Williams, Simple statistical gradient-following algorithms for connectionist reinforcement learning, Mach Learn **8**, 229 (1992).

[28] V. Mnih, K. Kavukcuoglu, D. Silver, A. Graves, I. Antonoglou, D. Wierstra, and M. Riedmiller, *Playing Atari with Deep Reinforcement Learning*, arXiv:1312.5602.

[29] J. Schulman, F. Wolski, P. Dhariwal, A. Radford, and O. Klimov, *Proximal Policy Optimization Algorithms*, arXiv:1707.06347.

[30] H. van Hasselt, A. Guez, and D. Silver, *Deep Reinforcement Learning with Double Q-Learning*, arXiv:1509.06461.

[31] J. J. M. Verbaarschot, H. A. Weidenmüller, and M. R. Zirnbauer, Grassmann integration in stochastic quantum physics: The case of compound-nucleus scattering, Physics Reports **129**, 367 (1985).

[32] I. Rotter, Effective Hamiltonian and unitarity of the S matrix, Phys. Rev. E **68**, 016211 (2003).

[33] Y. V. Fyodorov and H.-J. Sommers, Statistics of resonance poles, phase shifts and time delays in quantum chaotic scattering: Random matrix approach for systems with broken time-reversal invariance, Journal of Mathematical Physics **38**, 1918 (1997).

[34] A. F. Oskooi, D. Roundy, M. Ibanescu, P. Bermel, J. D. Joannopoulos, and S. G. Johnson, Meep: A flexible free-software package for electromagnetic simulations by the FDTD method, Computer Physics Communications **181**, 687 (2010).

[35] T. P. Dussauge, W. J. Sung, O. J. Pinon Fischer, and D. N. Mavris, A reinforcement learning approach to airfoil shape optimization, Sci Rep **13**, 9753 (2023).

[36] L. Feng, Y.-L. Xu, W. S. Fegadolli, M.-H. Lu, J. E. B. Oliveira, V. R. Almeida, Y.-F. Chen, and A. Scherer, Experimental demonstration of a unidirectional reflectionless parity-time metamaterial at optical frequencies, Nature Mater **12**, 108 (2013).

[37] M. Davy, Z. Shi, and A. Z. Genack, Focusing through random media: Eigenchannel participation number and intensity correlation, Phys. Rev. B **85**, 035105 (2012).

[38] X. Cheng and A. Z. Genack, Focusing and energy deposition inside random media, Optics Letters, Vol. 39, Issue 21, Pp. 6324-6327 (2014).

[39] M. Davy, Z. Shi, J. Wang, X. Cheng, and A. Z. Genack, Transmission Eigenchannels and the Densities of States of Random Media, Phys. Rev. Lett. **114**, 033901 (2015).